# Statistical Issues and Recommendations for Clinical Trials Conducted During the COVID-19 Pandemic

R. Daniel Meyer[a]*, Bohdana Ratitch[b], Marcel Wolbers[c], Olga Marchenko[d], Hui Quan[e], Daniel Li[f], Chrissie Fletcher[g], Xin Li[c], David Wright[i], Yue Shentu[j], Stefan Englert[k], Wei Shen[m], Jyotirmoy Dey[n], Thomas Liu[o], Ming Zhou[f], Norman Bohidar[r], Peng-Liang Zhao[e], Michael Hale[t]

[a]*Pfizer Inc., Groton, CT, USA;* [b]*Bayer, Montreal, Canada;* [c]*F. Hoffmann-La Roche Ltd, Basel, Switzerland;* [d]*Bayer, Whippany, NJ, USA ;* [e]*Sanofi, Bridgewater, NJ ;* [f]*Bristol Myers Squibb, NJ, USA ;* [g]*GlaxoSmithKline, Stevenage, Hertfordshire, UK, ;* [i]*AstraZeneca, Cambridge, Cambridgeshire, UK ;* [j]*Merck & Co., Inc., Rahway, NJ, USA;* [k]*Abbvie Deutschland GmbH & Co KG, Ludwigshafen, Germany ;* [m]*Eli Lilly and Company, Indianapolis, IN;* [n]*Abbvie Inc., North Chicago, IL, USA ;* [o]*Amgen Inc., Thousand Oaks, CA, USA;* [r]*Johnson&Johnson, Spring House, PA, ;* [t]*Takeda, Cambridge, MA, USA;*

**ABSTRACT**

The COVID-19 pandemic has had and continues to have major impacts on planned and ongoing clinical trials. Its effects on trial data create multiple potential statistical issues. The scale of impact is unprecedented, but when viewed individually, many of the issues are well defined and feasible to address. A number of strategies and recommendations are put forward to assess and address issues related to estimands, missing data, validity and modifications of statistical analysis methods, need for additional analyses, ability to meet objectives and overall trial interpretability.

Subject classification codes: NA

## 1. Introduction

The COVID-19 outbreak emerging in China in December 2019 quickly became a global pandemic as declared by the World Health Organization in March 2020. As of today, still only a few months into the pandemic, the disease and public health control measures are having very substantial impact on clinical trials globally. Quarantines, site restrictions, travel restrictions affecting participants and site staff, COVID-19 infections of study participants, and interruptions to the supply chain for study medication have led to operational problems, including difficulties in adhering to study protocols. Trial sponsors have rapidly responded to this crisis, where the overriding primary concern has been to protect participant safety. Some trials have been halted or enrolment suspended in the interest of participant safety. For ongoing trials, sponsors have implemented a variety of mitigations to assure safety of participants and address operational issues.

The downstream effects of protocol deviations and trial conduct modifications lead to varying degrees of impacts on clinical trial data. The impacts, described in more detail in later sections, raise important statistical issues for the trial. In the extreme, trial integrity, interpretability, or the ability of the trial to meet its objectives could be compromised. Intermediate to that, planned statistical analyses may need to be revised or supplemented to provide a thorough and appropriate interpretation of trial results. This paper offers a spectrum of recommendations to address the issues related to study objectives, inference, and statistical analyses. The major categories of impacts and mitigations are summarized in Figure 1.

The issues we discuss here largely involve ongoing trials, started before but conducted during the pandemic for non-COVID-19 related therapies. Many of the issues and recommendations will also apply to new trials. Regulatory agencies have rapidly published guidance for clinical trial sponsors to address COVID-19 issues (FDA 2020, EMA 2020a, 2020b). The current paper is influenced by and expands upon these important guidance documents.

The paper is organized as follows. In Section 2 we describe overall trial impact assessment. Section 3 considers assessment of impacts on the trial through the estimand framework. Section 4 summarizes recommendations for revised and supplemental analyses that may be needed for the trial, including the likely mechanisms of missing data and the recommended statistical approaches to address missingness. Section 5 outlines additional considerations for trial-level impact. A summary of recommendations is given in Section 6.

Figure 1. Key dimensions of assessment, mitigations and documentation to address the COVID-19 impact

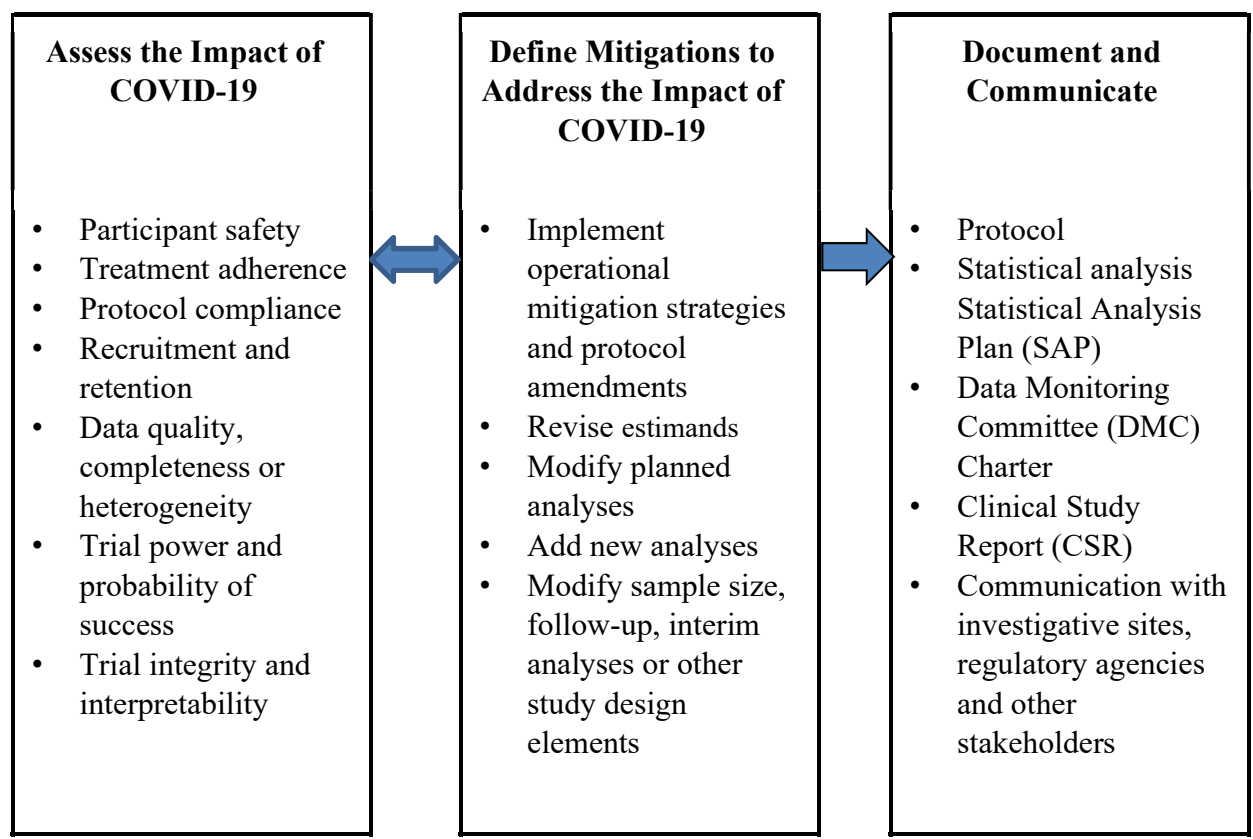

## 2. Pandemic-related Factors, Impacts, and Risk Assessment

Examples of pandemic-related impacts on trials that may lead to statistical issues are given in Table 1. Some of the impacts are directly caused by pandemic control measures or COVID-19 disease, while others result from operational modifications applied to address the direct impacts on trial conduct. The pandemic may have different effects on individual participants depending on when they enter a trial. The methods for summarizing pandemic effects on participants are discussed in Section 4.3.3.

Table 1. COVID-19-related factors and examples of their potential impacts on clinical trials.

| Factor | Example of Impact/ Risk |
|---|---|
| Quarantines, travel limitations, participant unable/unwilling to travel to site due to personal pandemic-related reasons, site closures or reduced availability of site staff | <ul><li>Missed or delayed visits and assessments</li><li>Inability to access study treatment</li><li>Loss to follow-up</li><li>Longer query response time</li><li>Different investigators / different measurement modalities</li><li>Delayed site monitoring</li><li>Delayed patient enrolment</li></ul> |
| Interruptions to supply chain of experimental drug and/or other medications | <ul><li>Missed dosing of study drugs</li><li>Changes in non-COVID-19 concomitant medications</li></ul> |
| Alternative administration of drug | <ul><li>Increased risk of dosing errors</li><li>Lack of equivalence of methods of administration</li></ul> |
| Alternative collection of specimens | <ul><li>Challenges in reconciliation and verification</li></ul> |
| Alternative data collection | <ul><li>Lack of exchangeability of methods</li></ul> |
| COVID-19 infection / treatment | <ul><li>Temporary / permanent interruption of study treatment and/or study participation</li><li>Potential effect on efficacy endpoints /estimands / safety</li><li>Interactions of COVID-19 concomitant medications with study drugs</li></ul> |

For trials impacted by the pandemic, assessing the change of the benefit/risk for participants is the first step in the decision-making process (FDA 2020). All recommendations in this article presuppose that appropriate steps have been taken to assure participant safety.

## 2.1. *Overall Risk Assessment*

Sponsors are advised to perform a risk assessment based on aggregated and blinded data to evaluate the likelihood of a trial to deliver interpretable results. It must start as a forward-looking assessment in anticipation of effects not yet seen but with some likelihood to occur. It should continue throughout the conduct of the study in light of the evolving situation and accumulating data, considering regional differences in the infection status and pandemic-control measures. All measures need to be taken to minimize sources of operational and statistical bias, especially regarding principal measures of efficacy and safety. Key elements are:

- Determine whether any trial modifications should be performed to protect participant safety and minimize risks to trial integrity;

- Identify all study participants who are affected by COVID-19 and understand how they are affected (e.g., treatment interruption, study discontinuation, missed visits) and what is the impact on the trial;

- Determine what additional information needs to be collected in the study database or in the form of input from study investigators in order to adequately monitor, document, and address pandemic-related issues (feasibility to obtain such information and its quality may vary and this needs to be considered as part of the risk factors);

- Understand reasons for treatment or study discontinuation and the impact on planned estimands and intercurrent events;

- Evaluate extent of missing data and specific reasons for missingness;

- Assess changes in enrollment and in study population over time;

- Evaluate the protocol-specified assumptions and the likelihood that the trial would be able to achieve its goals;

- Ideally, verify the usability of data captured from alternative methods (e.g., virtual audio or video visits) before implementing them. Such data may add more variability or not be interpretable;

- Determine any changes to planned analyses and analysis population definitions, or additional sensitivity analyses that need to be pre-specified prior to unblinding.

Based on the risk evaluations above, many sponsors have developed standardized metrics of trial operational status, such as rates of missed visits, discontinuations from treatment and study, protocol deviations, adverse events (AEs), to reinforce a consistent approach to risk monitoring and assessment. Such metrics are useful to identify trials that are more acutely impacted and to monitor the overall state of a portfolio of trials.

## 3. Implications and Mitigations for Estimands

The ICH E9(R1) Addendum (ICH, 2019) defines the estimand framework for ensuring that all aspects of study design, conduct, analysis, and interpretation align with the study objectives. It also provides a rigorous basis to discuss potential pandemic-related disruptions and to put them in the context of study objectives and design elements.

For an affected trial, the first major question is whether the primary objective, and therefore the primary estimand, should target the treatment effect without potentially confounding influences of COVID-19. We recommend that for most studies started before the pandemic, the original primary objective should be maintained as designed, implying a treatment effect that is not confounded by pandemic-related disruptions. (For new studies, this definition of treatment effect may also be reasonable, but depends on many aspects of the trial design.) This does not automatically imply a broad "hypothetical estimand" with the same hypothetical scenario for all possible pandemic-related intercurrent events (ICE). Confounding may need to be addressed in different ways for different types of ICEs depending on the study context. We discuss this in Section 3.1. Some changes in the original estimand definitions may, therefore, be needed to account for the pandemic-related disruptions. Sponsors may also consider additional estimands for exploratory purposes to characterize the treatment effect in specific conditions or sub-populations that emerged during the COVID-19 pandemic.

In studies where estimand(s) have been explicitly defined in the original protocol, sponsors should examine the effect of pandemic-related disruptions on estimand attributes and modify them as appropriate. The necessary revisions may vary under different risk mitigation scenarios, such as continuing study enrolment as planned versus temporarily suspending enrolment. For studies without explicitly defined estimands, this section may be useful to guide thinking about pandemic-related impacts on study objectives, conduct, inference, and statistical analyses.

Below we review each attribute of the estimand definition and highlight important pandemic-related considerations. In keeping with the ICH E9(R1) guideline, we make a distinction between intercurrent events and missing data handling. Intercurrent events are discussed in this section as one of five attributes used to define an estimand. Missed assessments that would have been meaningful for a given estimand if collected, while the study treatment continues as planned, should not be considered as ICEs. Such situations lead to missing data and should be considered as part of the missing data strategies in the context of an estimator, albeit in alignment with the estimand (see Section 4.2).

Our discussion is mainly geared towards considerations for the primary efficacy estimands, but similar logic can be applied to other study estimands.

### 3.1. Handling of Intercurrent Events

Strategies for handling non-pandemic related ICEs should remain unchanged. Here we discuss handling of pandemic-related ICEs only. The estimand framework allows for different strategies to be used for different types of ICEs and such estimands will likely be the most appropriate in the current context.

ICEs should be considered pandemic-related if they occur as a result of pandemic-related factors and are not attributed to other non-pandemic related reasons, e.g., treatment discontinuation due to lack of efficacy or a toxicity. Pandemic-related ICEs of importance should first be categorized in terms of their impact on study treatment adherence (e.g., study treatment discontinuation) or ability to ascertain the target outcomes (e.g., death). These ICEs should then be further categorized according to pandemic-related factors in order to choose the appropriate strategies to handle them. The relevant factors are study treatment accessibility, participant's COVID-19 infection condition, and participant's concomitant treatment for COVID-19 in case of infection. The key attributes that, when taken together, describe each pandemic-related ICE are summarized in Table 2. For example, premature study treatment discontinuation (listed in row 1 of Table 2) would only be considered pandemic-related if it is caused by one or more pandemic-related factors (listed in rows 2-4 of Table 2) and not attributed to other factors such as lack of efficacy or study treatment toxicity. (The list is not exhaustive and additional categories may need to be considered in some studies.) The combination of factors and therefore distinct varieties of ICEs is potentially large, but it may be decided that all or most ICEs can be handled in a similar manner.

Table 2. Attributes of pandemic-related intercurrent events.

| Participant's Adherence to Study Treatment | <ul><li>Study treatment permanently discontinued, and no new treatment started for disease under study;</li><li>Study treatment permanently discontinued and switched to an alternative therapy for disease under study;</li><li>Study treatment temporarily interrupted, or compliance significantly reduced without changes in concomitant therapy for disease under study;</li><li>Study treatment temporarily interrupted, or compliance significantly reduced with changes in the concomitant therapy for disease under study, e.g., start of rescue medications.</li></ul> |
|---|---|
| Study Treatment Accessibility | <ul><li>Study/region-wise drug supply interruption;</li><li>Site unavailable to administer/dispense study treatment;</li><li>Study treatment available at site but participant is unable/unwilling to get study treatment due to personal pandemic-related reasons.</li></ul> |
| Participant's COVID-19 Infection Condition | <ul><li>Participant positive for COVID-19 and alive;</li><li>Participant deceased due to COVID-19*;</li><li>Participant with a suspected COVID-19 infection (may be distinguished further by presence or absence of symptoms).</li><li>Participant with an exacerbation of underlying health issues due to reduced healthcare access.</li></ul> |

| **Participant's COVID-19 Concomitant Treatment(s) *** | • Participants treated for COVID-19 (pharmacologically, oxygen, etc.); <br> • Hospitalized, not in intensive care; <br> • Admitted to intensive care. |
|---|---|

\* COVID-19 related deaths and initiation of treatment for COVID-19 infections may also be considered as ICEs if they occur after the completion of study treatment or after other non-pandemic-related ICEs and before the time point associated with the endpoint of interest.

Certain types of non-adherence to study treatment may not normally be considered as ICEs but may need to be in the context of the pandemic. For example, an ICE of significantly reduced compliance or temporary treatment interruption may not have been anticipated at study design but could be now if considered likely due to pandemic-related disruptions. For some studies with significant pandemic-related treatment interruptions, the minimal duration of interruption expected to dilute the treatment effect could be defined. Different strategies can be used for interruptions exceeding this duration as opposed to shorter interruptions. Sensitivity analyses can be used to assess robustness of inference to the choice of cut-off. For time-to-event endpoints, it is tempting to define the minimally acceptable level of drug compliance on a participant level according to the observed exposure to study drug from the first dose of study drug until the event or censoring and exclude participants without a minimally acceptable level of compliance. However, such an approach could introduce immortal time bias and should therefore be avoided (van Walraven et al, 2004).

A special consideration may be warranted for participants receiving experimental treatment for COVID-19 regardless of whether they remain on study treatment. Also, in studies where mortality was not originally expected, death due to COVID-19 should be considered as a potential ICE.

Most of the ICE types listed against the "Participant's Adherence to Study Treatment" attribute in Table 2 (e.g., study treatment discontinuation) due to non-pandemic related reasons are likely addressed in the primary estimand prior to the pandemic. We recommend starting with an examination of whether the original strategy is justifiable when these ICEs occur due to the pandemic. If not, a different strategy should be chosen. We outline some high-level considerations in this respect.

- *Treatment policy strategy*, in which ICEs are considered irrelevant in defining the treatment effect, will typically not be of scientific interest for most pandemic-related ICEs because the conclusions would not generalize in the absence of the pandemic. For example, the treatment effect estimated under the treatment policy for premature treatment discontinuations caused by pandemic-related disruptions will reflect the effect of a regimen where discontinuations and changes in therapy occur due to pandemic-related factors (e.g., disruptions in study drug supply) which would not be aligned with the primary study objective. Initiation of treatment for COVID-19 infection after an earlier non-pandemic-related ICE that was planned to be handled by the treatment policy strategy but prior to observation of an efficacy or safety endpoint will also need to be considered carefully and cannot simply be deemed irrelevant. Under the treatment policy strategy, the estimated treatment effect may reflect the effects of infection and its treatment, which are presumably not of interest for the primary objective.

  A decision to use a treatment policy approach for pandemic-related ICEs may be justifiable if the percentage of participants with such events is low and this strategy was planned for similar non-pandemic related ICEs. This strategy may also be considered for handling ICEs corresponding to relatively short treatment interruptions. The treatment policy strategy should be avoided for pandemic-related ICEs of premature study treatment discontinuation in non-inferiority and equivalence studies, as similarity between treatment groups may artificially increase with the number of such events. Similar considerations also apply to the composite strategy.

- *Composite strategies*, in which ICEs are incorporated into the definition of the outcome variable, are unlikely to be appropriate for most pandemic-related ICEs. For example, study treatment discontinuation due to pandemic-related disruptions should not be counted as treatment failure in the same way as discontinuation due to lack of efficacy or adverse reactions. A more nuanced consideration may be needed in studies of respiratory conditions, where COVID-19 complications may be considered with a composite strategy as a form of unfavourable outcome. (See also a discussion on COVID-19 related deaths further below.)

- *Principal stratification strategy* stratifying on a COVID-19 related event (e.g., serious complications or death due to COVID-19) is unlikely to be of interest for the primary estimand because it would limit conclusions to a sub-population of participants defined based on factors not relevant in the context of the future clinical practice.

- *While-on-treatment strategy* may continue to be appropriate under certain conditions if it was originally planned for non-pandemic-related ICEs. This strategy is typically justifiable when treatment duration is not relevant for establishing treatment effect (e.g., treatment of pain in palliative care), but certain conditions may need to be considered, such as a minimum treatment duration required to reliably measure treatment outcomes.

- *Hypothetical strategy,* in which the interest is in the treatment effect if the ICE did not occur, is a natural choice for most pandemic-related ICEs. This would especially apply to ICEs of study treatment disruption for pandemic reasons. For such participants, the hypothetical scenario where they would continue in the study in the same way as similar participants with an undisrupted access to treatment is reasonable. It is not necessary to assume a hypothetical scenario where such participants would fully adhere to the study treatment through the end of the study. Rather, a hypothetical scenario may include a mixture of cases who adhere to the study treatment and those who don't adhere for non-pandemic reasons. Discussions with regulatory agencies may be helpful to reach an agreement on the details prior to the final study unblinding.

  Although estimation methods are not part of the estimand consideration, the ability to estimate treatment effects in a robust manner under a hypothetical strategy based on available data should not be taken for granted and should be assessed as the estimand definition is finalized. This aspect should be part of the overall risk assessment and decisions on the choice of the mitigation strategies.

The discussion above highlights the need to capture the information associated with pandemic-related factors, such as those listed in Table 2. This can be done either through designated fields in the Case Report Form (CRF) or through a detailed and structured capture of protocol deviations.

An ICE of death due to COVID-19 requires careful consideration and the appropriate strategy depends on the disease under study and clinical endpoint. In disease areas with minimal mortality where death is not a component of the endpoint, a hypothetical strategy for deaths related to COVID-19 infections may be recommended. For studies in more severe diseases where death is part of the endpoint, it is inevitable that more than one estimand will be of interest when evaluating the benefit of treatment for regulatory purposes. A pragmatic approach which includes COVID-19-related deaths in the outcome, i.e., which uses a composite strategy, is suitable if the number of COVID-19-related deaths is low or if there is a desire to reflect the impact of the pandemic in the treatment effect estimate. (See also the related section on competing risks analyses in Section 4.2.2.) Using a hypothetical strategy for deaths related to COVID-19 infections will be important in evaluating the benefit of treatment in the absence of COVID-19 (for example, when the disease is eradicated or effective treatment options emerge). It is acknowledged that such trials frequently include elderly, frail, or immunocompromised participants and it may be difficult to adjudicate a death as caused by COVID-19 or whether the participant died with COVID-19.

While treatment policy, composite, and principal stratification strategies may not be of interest for the primary estimand, they may be of interest for supplementary estimands when there is a scientific rationale to investigate the study treatment either in sub-populations of participants stratified based on COVID-19 infection and outcomes and/or together with a concomitant use of treatments administered for COVID-19. For example, this may be of interest for studies in respiratory diseases or conditions suspected to be risk factors for COVID-19 complications. The relevance of such estimands will also depend on the evolution of this pandemic, whether the virus is eventually eradicated or persists like seasonal flu. In the latter case, the reality and clinical practice are still likely to be different from the current crisis management conditions as the society and clinical practice adapt to deal with a new disease.

### 3.2. Treatment Condition (Intervention) of Interest

In general, treatment condition of interest should remain the same as originally intended. However, operationally, the mode of treatment delivery may need to be changed due to pandemic-related reasons, e.g., treatment self-administered by the participant at home rather than in the clinic by a health-care professional. When such changes are feasible, they may be considered to reduce the frequency of visits to the clinic, and therefore reduce the risk of infection exposure.

Pandemic-related complications with study treatment adherence and concomitant medications should be considered as ICEs and handled with an appropriate strategy. The extent of such ICEs should be evaluated in terms of whether the treatment(s) received by participants during the study remain sufficiently representative of what was intended. This may include treatment interruptions, reduced compliance, and access to any background, rescue, and subsequent therapies planned to be covered by the treatment policy strategy.

### 3.3. Target Population

To align with the primary study objective, the target participant population should remain as originally planned, i.e., should not be altered simply due to the pandemic. Protocol amendments unrelated to the pandemic to further qualify the study population should still be possible. The trial inclusion/exclusion criteria should also remain largely unchanged relative to those that would be in place in absence of the pandemic, except for the possible exclusion of active COVID-19 infections.

### 3.4. Variable (Endpoint)

The clinical endpoint should generally remain as originally planned. In cases where alternative measurement modalities may be necessary during the pandemic, for example, central labs vs. local labs, remote assessments of questionnaires instead of in-clinic, etc., it must be assured that clinical endpoint measurement is not compromised and potential effects on endpoint variability should be assessed (see Section 4.3.2). In cases where pandemic-related ICEs, such as COVID-19 deaths, are handled using the composite strategy, the definition of the endpoint may need to be adjusted. In cases of numerous delays between randomization and start of treatment, where the endpoint is defined relative to the date of randomization (e.g., in time-to-event endpoints), consideration may be given to redefine the endpoint start date to start of treatment. However, in the context of open-label studies this may not be advisable.

### 3.5. Population-level Summary of Treatment Effect

Population-level summary describing outcomes for each treatment and comparison between treatments should remain unchanged, in general. In rare situations, a summary measure may need to be changed, for example, if the originally planned endpoint is numeric and a composite strategy is used for COVID-19 deaths to rank them worse than any value in survived participants. In this case, a summary measure may be changed from mean to median.

Another example could be a hazard ratio (HR) from a Cox proportional hazard regression, a summary measure of treatment effect commonly used for trials with time-to-event endpoints. If the assumption of proportional hazards is not satisfied, the estimated HR depends on the specific censoring pattern observed in the trial, which is influenced both by participant accrual and dropout patterns (Rufibach, 2019). External validity and interpretability of the HR needs to be carefully considered if censoring patterns are affected during the pandemic in ways that are not representative of non-pandemic conditions and if additional pandemic-related censoring depending on covariates such as the participant's age or comorbidities are observed. Similarly, the validity of the Log-rank test relies on the assumptions that the survival probabilities are the same for participants recruited early and late in the trial and that the events happened at recorded times. Such assumptions may need to be assessed. Supportive estimands with alternative summary measures could be considered (see e.g., Boyd et al., 2012; Nguyen and Gillen, 2012; Mao et al., 2018).

## 4. Implications and Mitigations for Analysis

Planned statistical analyses may need to be modified due to effects of the pandemic on trials. Additional sensitivity and supplementary analyses may be needed to properly understand and characterize the treatment effect. Depending on the trial, modifications in planned analyses may range from relatively minor, e.g., for trials with relatively low impact, to major, e.g. in settings where study drug administration and visits are severely disrupted by the pandemic. A general summary of analysis considerations is provided in Table 3 and detailed discussions are presented in subsequent sections.

All planned modifications and additional analyses should be documented in the SAP prior to data unblinding and in the clinical study report. Additional post hoc exploratory analyses may also be necessary after study unblinding to fully document the impact of the pandemic and characterize the treatment effect in this context.

Table 3: Summary of analysis considerations

| **Review of planned analyses** | • Review all planned main and sensitivity analyses to ensure alignment with the revised estimand(s).<br>• Review / amend methods for handling of missing data, or censoring rules, to accommodate pandemic-related missingness. |
|---|---|
| **Summaries of pandemic impact** | • Summarize the occurrence of pandemic-related ICEs and protocol deviations.<br>• Summarize the number of missed or unusable assessments for all key endpoints.<br>• Summarize the number of assessments performed using alternative modalities.<br>• Summarize study population characteristics before and after pandemic onset. |
| **Additional sensitivity and supportive analyses** | • Plan additional analyses for sensitivity to pandemic-related missingness.<br>• Consider the need for additional, alternative summary measures of treatment effect.<br>• Consider exploring inclusion of additional auxiliary variables, interaction effects, and time-varying exogenous covariates in the analysis methods.<br>• Consider subgroup analyses based on subgroups defined by pandemic impact, e.g. primary endpoint visits before or after pandemic onset.<br>• Consider the need for evaluation of potential impact of alternative data collection modalities.<br>• Consider sources of data external to the trial, for example to justify use of alternative modalities.<br>• Plan for additional safety analyses. |

## 4.1. Considerations for Efficacy Analyses

All planned efficacy analyses should be re-assessed considering the guidance provided in Sections 3 and in terms of handling of pandemic-related missing data (see Section 4.2). The core analysis methodology should not change. However, depending on the revisions to the estimand, the strategies chosen for pandemic-related ICEs, and the handling of pandemic-related missing data, some changes to the planned analyses may be warranted. Additional analyses will frequently also be required to assess the impact of the pandemic disruption.

Special considerations may be needed for studies and endpoints where participant outcomes could be directly impacted by the pandemic, e.g., in respiratory diseases or quality of life endpoints.

In studies where enrolment is halted due to the pandemic, sponsors should compare the populations enrolled before and after the halt. More generally, shifts in the population of enrolled participants over the course of the pandemic should be evaluated. Baseline characteristics (including demographic, baseline disease characteristics, and relevant medical history) could be summarized by enrolment period to assess whether there are any relevant differences in the enrolled population relative to the pandemic time periods. Shifts could be associated with regional differences in rates of enrollment because start and stop of enrollment is likely to vary by country as pandemic measures are implemented or lifted.

### 4.2. Implications and Mitigations for Missing Data

Sponsors should make every effort to minimize the amount of missing data without compromising safety of participants and study personnel during the COVID-19 pandemic and placing undue burden on the healthcare system. Whenever feasible and safe for participants and sites, participants should be retained in the trial and assessments continued, with priority for the primary efficacy endpoint and safety endpoints, followed by the key secondary endpoints. Despite best efforts, sponsors should prepare for the possibility of increased amounts and/or distinct patterns of missing data.

In the framework of ICH E9(R1), an assessment or endpoint value is considered missing if it was planned to be collected and considered useful for a specific estimand but ended up being unavailable for analysis. In case of ICEs that are addressed by a hypothetical strategy, endpoint values are not directly observable under the hypothetical scenario. Such data are not missing in the sense of the ICH E9(R1) definition, however, they need to be modelled in the analysis, often using methods similar to those for handling of missing data. In the remainder of the paper, we will discuss methods for handling of missing data, and note that such methods can be useful for modelling unobserved data after ICEs, if the modelling assumptions align with the hypothetical scenarios chosen for addressing the corresponding ICEs.

*4.2.1. Assessing and Documenting Pandemic-related Missingness*

Sponsors should assess and summarize patterns (amount and reasons) of pandemic-related missing data in affected trials. Data may be missing because a) planned assessments could not be performed; b) collected data is deemed unusable for analysis, e.g., out-of-window; or c) data under a desired hypothetical scenario cannot be observed after an ICE (e.g., censored). Additionally, each pandemic-related missingness instance also has specific reasons and circumstances. At a high level, reasons for pandemic-related missing data could be structural (e.g., government enforced closures or sites stopping study-related activities) or they could be participant-specific (e.g., individual COVID-19 disease and complications or individual concerns for COVID-19). Table 4 outlines the aspects that together provide a comprehensive picture for assessing the impact of missing data and planning how to handle them in analysis. Sponsors should, therefore, capture such information in the clinical study database as much as possible. The last two rows of Table 4 reflect circumstances similar to those that are considered in the context of ICEs (see Table 2). Since missing data may occur both in the presence of ICEs (e.g., handled by a hypothetical strategy) and in the absence of ICEs (e.g., participant continues to adhere to study treatment but misses some assessments), we list them here as well.

Table 4. Attributes of pandemic-related missing data. Row 1 summarizes reasons of missing data, rows 2-4 summarize related conditions that contribute to those reasons.

| **Missing Endpoint Measurement** | <ul><li>Assessment missing due to a participant's premature discontinuation from the study overall for pandemic-related reasons;</li><li>Assessment missing due to missed study visits/procedures while a participant remains in the study (intermittent missing data);</li><li>Assessment delayed (out-of-window) and deemed unusable for an analysis;</li><li>A composite score (e.g., ACR20 in rheumatoid arthritis) cannot be calculated because some components are missing;</li><li>Assessment deemed to be influenced by pandemic-related factors and deemed unusable for a particular analysis because the interpretability of the results may be impacted (e.g., in assessments of quality of life, activity/functional scales, healthcare utilization, etc);</li><li>Recorded data cannot be properly verified or adjudicated due to COVID-19-related factors and deemed to be unreliable for analysis;</li><li>Assessment performed after an intercurrent event intended to be handled with a hypothetical strategy and collected data are deemed unusable for this estimand.</li></ul> |
|---|---|
| **Assessment Accessibility** | <ul><li>Site (facilities or staff) unavailable to perform study-related assessments;</li><li>Site/assessment procedure available but participant is unable/unwilling to get assessment done due to personal pandemic-related reasons.</li></ul> |
| **Participant's COVID-19 Infection Condition** | <ul><li>Participant positive for COVID-19 and alive;</li><li>Participant deceased due to COVID-19;</li><li>Participant without a known COVID-19 infection.</li></ul> |
| **Participant's COVID-19 Concomitant Treatment(s)** | <ul><li>Participants treated for COVID-19 (pharmacologically, oxygen, etc.);</li><li>Hospitalized, not in intensive care;</li><li>Admitted to intensive care.</li></ul> |

Sponsors should also consider a potential for under-reporting of symptoms and AEs during the pandemic due to missed study visits or altered assessment modalities, e.g., a telephone follow-up instead of physical exam. (See Section 4.4.)

Sponsors should also consider reporting patterns of missingness along several dimensions: over time (in terms of study visits as well as periods before and after the start of pandemic disruptions), with respect to certain demographic and baseline disease characteristics, as well as co-morbidities considered to be potential risk factors associated with COVID-19 infection or outcomes, and across geographic regions.

Blinded summaries of missing data patterns prior to study unblinding may inform the choice of missing data handling strategies. It may be useful to compare missing data patterns from the current studies with similar historical studies, especially with respect to missingness in subgroups.

After study unblinding (for final or unblinded DMC analyses), missing data patterns should be summarized overall and by treatment arm. Although in most cases pandemic-related missingness, especially structural missingness, would not be expected to differ between treatment arms, such a possibility should not be ruled out. In special circumstances, such as in an open-label study, missing visits may be related to treatment if the experimental treatment must be administered at the site while the control treatment could be administered at home, there may be more missing assessments in the control group. This may result in biased treatment effect estimates if mitigating strategies are not implemented. This could also be the case for a single cohort trial using an external control.

### 4.2.2. *Handling of Missing Data in Main Analyses*

Sponsors should generally maintain the same approaches for handling of non-pandemic-related missing data as originally planned in the protocol and SAP. For pandemic-related missingness, appropriate strategies will need to be identified in the context of each estimand and analysis method. Which strategy is most appropriate should be considered in light of the underlying context and reasons for missingness as shown in Table 4 and in alignment with the estimand for which the analysis is performed. Three cases are described.

(1) When data are missing without the participant having an ICE, i.e., participant continues to adhere to study treatment but has some endpoint values missing: the missing data modelling should be based on clinically plausible assumptions of what the missing values could have been given the fact the participant continues to adhere to study treatment and the participant's observed data.

(2) When data are modelled in presence of an ICE: the strategy defined in the estimand for addressing that ICE should be considered.

(3) When modelling outcomes after an ICE addressed by a hypothetical strategy: the assumptions should reflect what the unobserved outcome could have been under the specific hypothetical scenario specified in the estimand and be clinically plausible in the context surrounding the ICE. For example, unobserved data can be modelled differently for ICEs of study treatment discontinuation depending on whether the participants suffer from COVID-19 infection complications or not, as having such complications may indicate a different overall health state.

Imputation methods or methods which implicitly deal with missing values recommended in the current literature and by regulatory agencies should, in most cases, provide an adequate selection of tools to deal with pandemic-related missingness (see e.g., Molenberghs and Kenward, 2007; NRC, 2010; O'Kelly and Ratitch, 2014, Mallinckrodt et al., 2019).

Methods for dealing with missing data are often categorized based on the type of assumptions that can be made with respect to the missingness mechanism. Using

Molenberghs and Kenward (2007)'s classification of missingness mechanisms aligned with longitudinal trials with missing data, data are missing completely at random (MCAR) if the probability of missingness is independent of all participant-related factors or, conditional upon appropriate pre-randomization covariates, the probability of missingness does not depend on either the observed or unobserved outcomes. (We note that in the framework of Little and Rubin (2002), MCAR is defined as independent of any observed or unobserved factors. This definition was subsequently generalized to encompass dependence on pre-randomization covariates, also referred to as covariate-dependent missingness, and MCAR is now used in the literature in both cases.) Some types of pandemic-related missingness may be considered MCAR, e.g., if it is due to a site suspending all activities related to clinical trials. Consideration may be given to whether such participants should be excluded from the primary analysis set depending on the amount of data collected prior to or after the pandemic. For example, when all (most) data are missing for some participants, imputing their data based on a model from participants with available data would not add any new information to the inference, while excluding such participants is unlikely to introduce bias. When participants have data only for early visits before the expected treatment effect onset and the rest of the data are MCAR, then including such participants in the analysis set may not add information for inference about treatment effect while adding uncertainty due to missing data.

Data are MAR if, conditional upon appropriate (pre-randomization) covariates and observed outcomes (e.g., before participant discontinued from the study), the probability of missingness does not depend on unobserved outcomes. If relevant site-specific and participant-specific information related to missingness is collected during the study, most of the pandemic-related missingness can be considered MCAR or MAR.

The definitions of MCAR and MAR mechanisms are based on conditional independence of missing data given a set of covariates and observed outcomes that explain missingness. The factors explaining pandemic-related missingness may include additional covariates and, in the case of MAR, post-randomization outcomes. For example, missingness during the pandemic may depend on additional baseline characteristics, e.g., age and co-morbidities, as well as post-randomization pandemic-related outcomes, such as COVID-19 infection complications.

In the case of covariate-dependent missingness, regression adjustment for the appropriate baseline covariates is sufficient for correct inference, though this complicates the analysis model and the interpretation of the treatment effect for models such as logistic or Cox regression where conditional and marginal treatment effect estimates do not agree. Under MCAR and MAR, some modelling frameworks such as direct likelihood, e.g., mixed models for repeated measures (MMRM), can take advantage of separability between parameter spaces of the distribution of the outcome and that of missingness. In this case, missingness can be considered ignorable (Molenberghs and Kenward, 2007), and the factors related only to missingness do not need to be included in the inference about marginal effects of treatment on outcome. This does not, however, apply to all inferential frameworks. Multiple imputation (MI) methodology (Rubin, 1987) may be helpful in this respect as it allows inclusion of auxiliary variables (both pre- and post-randomization) in the imputation model while utilizing the previously planned analysis model. Multiple imputation with auxiliary variables may be used for various types of endpoints, including continuous, binary, count, and time-to-event and coupled with various inferential methods in the analysis step. The use of MI with Rubin's rule for combining inferences from multiple imputed datasets may introduce some inefficiencies and impact study power, although some alternatives exist (see, e.g., Schomaker and Heumann, 2018; Hughes et al., 2016).

For implementing a hypothetical strategy for COVID-19 related ICEs in the context of time-to-event endpoints, regression models (e.g., Cox regression) adjusted for relevant baseline covariates (in the MCAR setting) or multiple imputation (in the MAR setting) is recommended. Competing risks analyses which treat pandemic-related ICEs that fully or partially censor the outcome, e.g., COVID-19 related deaths, as competing events are not compatible with a hypothetical strategy. A further complication in the interpretation of competing risks analyses in this context is that participants are not at risk for the competing event from their time origin (e.g., randomization or start of treatment) onwards but only during the pandemic which is not experienced synchronously across the trial cohort. This compromises the validity of common competing risks analyses and prevents interpretation of results from such analyses to be generalized to the population.

The implication of assuming specific missingness mechanism is that missing outcomes can be modelled using observed pre-randomization covariates or covariates and post-randomization outcomes (MAR) from other "similar" participants, conditional on the observed data. For pandemic-related missingness, it is important to evaluate whether there are sufficient observed data from "similar" participants to perform such modelling, even if factors leading to missingness are known and collected. For example, if missingness and endpoint depend on age, and few older participants have available endpoint data, it may not be appropriate to model missing data of older participants using a model obtained primarily from available data of younger participants (possibly resulting in unreliable extrapolation).

Similarly, severe complications of a COVID-19 infection may be due to these participants having a worse health state overall and modelling their outcomes based on data from participants without such complications may not be justifiable. Additional assumptions about participants with missing data versus those with observed data may need to be postulated and justified, perhaps based on historical data. This is where the consideration of pandemic-related factors surrounding missingness, such as those mentioned in Table 4, is important.

Pandemic-related missing data may need to be considered MNAR if missingness and study outcomes depend on COVID-19 related risk factors, treatment, and infection status but such data are not collected. The missingness mechanism may also be MNAR if it depends on unobserved study outcomes. In the context of the pandemic, it may arise when participants with milder disease or lower treatment response are more inclined to discontinue the study or treatment *and* if their outcomes and reasons for discontinuation are not documented before discontinuation. Analysis under MNAR requires additional unverifiable assumptions but may be avoided through collection of relevant data. Analysis of sensitivity to departures from MAR assumptions should be considered, for example, models assuming plausible MNAR mechanisms (see e.g., Carpenter et al., 2013; Mallinckrodt et al., 2020).

Modifications to planned main analyses needed to handle pandemic-related missing data should be specified in the SAP prior to study unblinding. See Section 4.3.1 for a discussion of sensitivity analyses with respect to missing data.

### *4.3. Considerations for Sensitivity and Supplementary Analysis*

Additional sensitivity and supplementary analyses will frequently be required to assess the impact of pandemic-related disruptions on the trial. For non-pandemic-related events and missing data, the originally planned sensitivity analyses should be performed, but simply applying the pre-planned sensitivity analyses to both pandemic and non-pandemic ICEs and missing data may be problematic for three reasons. First, the objective to estimate treatment effects in the absence of a COVID-19 pandemic may mandate different strategies for pandemic-related and unrelated events. Second, as discussed in Section 4.2.2, an MCAR or MAR assumption is frequently plausible for COVID-19-related events whereas this may not be the case for other missing data. Third, sensitivity analyses to missing data could become excessively conservative if the amount of pandemic-related missing data is large. While a relatively large proportion of missing data could normally be indicative of issues in trial design and execution leading to greater uncertainty in trial results, that premise is tenuous when an excess of missing data is attributed to the pandemic.

Subgroup analyses for primary and secondary endpoints by enrolment or pandemic (see Section 4.3.3) period are recommended. If subgroup analyses are indicative of potential treatment effect heterogeneity, the potential for this to be rooted in regional differences should be considered. Issues of multiregional clinical trials as described in ICH (2017) may be magnified by the pandemic. In addition, dynamic period-dependent treatment effects could also be assessed in an exploratory fashion. For example, in models for longitudinal and time-to-event endpoints, one could include interaction terms between the treatment assignment and the exogenous time-varying covariate describing the patients' dynamic status during follow-up. However, results from such analyses may be nontrivial to interpret and generalize.

### 4.3.1. Sensitivity to Delayed Assessments and Missing Data

More liberal visit time-windows may be appropriate for visits depending on the specific trial. Sensitivity analyses should assess the robustness of results to out-of-schedule visits by either including them or treating them as missing data.

A tipping point analysis (Ratitch et al., 2013) may be used to assess how severe a departure from the missing data assumptions made by the main estimator must be to overturn the conclusions from the primary analysis. Tipping point adjustments may vary between the pandemic vs. non-pandemic missingness and by reason of missingness. For example, one could tip missing data due to pandemic-unrelated missing data but use standard MAR imputation for pandemic-related missing data. Historical data may be useful to put in context the plausibility of the assumptions and tipping point adjustments.

When main analysis relies on imputation techniques, sensitivity analyses can be done by using an extended set of variables in the imputation algorithm.

When dealing with missing data in a context of intercurrent events handled with the hypothetical strategy, one could vary the assumed probability that the participant would have adhered to treatment through the end of the study vs that they would have had other non-pandemic-related events and impute their outcomes accordingly.

For binary responder analysis, it is not uncommon to treat participants who have missing assessments as non-responders when a proportion of such cases is very small. For pandemic-related delayed and missed assessments, especially those occurring in absence of ICEs, it may be preferable to use a hypothetical strategy based on a MAR assumption instead. However, anon-responder imputation for both non-pandemic and pandemic-related missing response assessments could be reported as a supplementary analysis. For time-to-event endpoints, we recommend the usage of interval-censored methods to account for cases where the event of interest is known to have occurred during a period of missing or delayed visits but the time to event is not precisely known in sensitivity analyses (Bogaerts, Komarek, and Lesaffre, 2017).

### 4.3.2. Sensitivity to Alternatives to Protocol-specified Study Data Collection

Alternative measurements of endpoints may be necessary during the pandemic period. A careful study-specific assessment is necessary to judge whether these alternative measurements are exchangeable with standard protocol assessments. Ideally, exchangeability is established at the time of implementation based on information external to the trial. If not, blinded data analyses can support this, e.g. comparisons of the distribution of alternative measurements to the original measurement. However, if the validity of the new instrument has not been established previously, it will be challenging to rigorously demonstrate equivalence using data from the trial alone.

If exchangeability can be established or assumed, the main estimator could include data from both the original and the alternative data collection. The sensitivity analysis would then include data collected according to the original protocol only and treat other data as missing. If the validity of the exchangeability assumption is uncertain, the opposite approach can be taken. Modelling the interaction between treatment and assessment method can be undertaken as an alternate sensitivity analysis.

### 4.3.3. Challenges in Understanding the General Pandemic Effect on Trial Outcomes

It will be important to understand the pandemic effect on trial outcomes. There are several schools of thought on how this could be done. Previous sections in this paper have described the need for collecting data describing how events such as missed visits and treatment interruptions can be attributed to the pandemic. These data are incorporated into definitions of pandemic-related ICEs and missing data, and the strategies for handling those in the analysis. Statistical analyses of trial data will then be properly adjusted for pandemic effects. In many ways this is the ideal approach as it incorporates what is known for each participant directly into the analysis, and in a way that is very well understood. This is a standard approach to adjusting statistical analysis for inevitable perturbations in clinical trials. This approach has the disadvantage of needing to collect detailed data on pandemic-relatedness, which may not be feasible in some circumstances.

Another approach involves the use of pandemic time periods defined external to the trial database (e.g., to define pre-/during/post-pandemic phases as described in EMA/CHMP 2020b). This approach requires the accurate and precise definition of pandemic periods. This is simpler to apply in single-country studies where the impact of the pandemic and local containment measures may be relatively homogeneous across participants. However, even for a single country, the pandemic may evolve in a gradual fashion, complicating the definition of pandemic start and stop dates, and the impact of the pandemic on study participants will likely not be homogenous. Moreover, there may be several waves of infection outbreaks. The definition could prove even more challenging to implement in global trials because the start and stop dates of these periods and the impact of the pandemic on study participants may well vary by region. In practice, a standardized and pragmatic definition based on regionally reported numbers of COVID-19 cases and deaths over time and/or start date and stop dates of local containment measures will likely be required. Once pandemic periods have been defined, time-varying indicator variables for visits occurring during different pandemic periods could be incorporated as appropriate in statistical models or in ICE definitions, as discussed previously. When this approach is used, the rationale for defining the pandemic start and stop dates should be documented. The situation is evolving rapidly and at this is point, it is not possible to provide definitive recommendation on the definition, implementation, and interpretation of these pandemic periods.

In a third approach to generally assessing pandemic effects, each participant in the trial could be categorized according to the extent of pandemic impact on their treatment and assessments collected in the study database (details of protocol deviation, ICEs, missing assessments, pandemic-related reasons for discontinuation etc.). For trials with a fixed follow-up duration and minimal loss to follow-up, the categorization could be integrated in standard analyses, for example in defining subgroups for standard subgroup analysis.

There is insufficient evidence currently to favour a single approach to this issue. Sponsors are preparing to do at least the first two approaches, as these have been the subject of regulatory guidance. Until we see how they play out and how the pandemic evolves, etc., it's sensible to consider multiple methods of summarizing pandemic effects.

### *4.4.    Considerations for Safety Analyses*

Standard safety summaries will include all AEs as usual. However, additional separate analyses may be needed for events associated with COVID-19 infections and unassociated events, respectively, to fully understand the safety profile. The determination whether AEs and, particularly, deaths are COVID-19 related should be made during trial monitoring before data unblinding to avoid bias.

In many situations, safety reporting will remain unchanged. However, disruptions due to the COVID-19 pandemic may lead to increases in treatment interruptions, discontinuations and study withdrawals as well as the occurrence of COVID-19 infections and deaths. Hence, the estimands framework outlined in Section 3 could also be useful for safety analyses and we refer to (Unkel et al, 2019) for a general discussion of estimands as well as time-to-event and competing risks analyses for safety. Trials that require physical visits to adequately assess safety of the intervention will need to have maintained a schedule of physical visits to satisfy the requirement. Generally addressing the potential for bias in collection of AE data is beyond the scope of this paper. Exposure considering the compliance rate- or follow-up-adjusted analyses could be done, e.g., comparing the adjusted rates before and during the pandemic or with historical data. We don't have other methodological recommendations at this time, and more research is needed.

## 5. Study-Level Issues and Mitigations

The cumulative impact of missing data and revised statistical models discussed in the previous sections contribute to an overall study-level impact. The cumulative effect could alter the likelihood of meeting trial objectives, or even the interpretability of the trial results.

### *5.1. Assessing Impact of Missing Data and Study Drug Interruptions / Discontinuations*

Sponsors should assess potential impact of missing data on several aspects and it may be important to reach agreement with regulatory agencies on some of these questions:

- Feasibility of planned estimation methods given the data expected to be available;
- Potential for bias in treatment effect estimation if there are important differences in missingness patterns across treatment arms;
- Study power for the primary and key secondary objectives;
- Interpretability of study findings;

- Adequacy of safety database due to potential reduction in total drug exposure time and potential for underreporting of AEs;
- Adequacy of regional evidence required for regulatory submissions.

### *5.2. Power and Probability of Success*

As discussed in the previous sections, COVID-19 related factors impact trial data in many ways with consequences for power of the study, probability of success, sample size or other aspects of the trial design.

Quantifying the potential effects of the various pandemic factors on trial results can be done through clinical trial simulations. The simulation models will depend on the factors used in the original trial design, and incorporate adjustments to estimands, missing data handling and analysis methods as discussed in Sections 3 and 4. To maintain trial integrity, the simulations should be informed only by blinded data from the study and the assumed values for the design parameters from external sources. Variability and treatment effect estimates may be modified from their original values used in trial design. Trial properties such as power and probability of success can be updated accordingly. Sample size adjustment can be considered based on the simulation outcomes, or more extensive modifications of the trial design also may be considered such as change in the primary endpoint, analysis method, or addition of interim analysis with associated adaptation (Posch and Proschan 2012; Kieser and Friede 2003; Muller and Schaefer 2004). Such changes are challenging, and should be discussed with regulatory agencies, but can be considered if trial integrity is maintained.

For some trials, it may not be feasible to increase sample size and the trial will fall short of enrollment target. Given the extraordinary circumstances, we advocate more flexibility to consider methods for quantifying evidence across multiple trials and sources, including use of historical control arm data and real-world data, although sources and methodology for selection of such data would need to be planned and agreed with regulatory agencies in advance. If the observed treatment effect after data unblinding is meaningful but does not meet the statistical criterion due to COVID-19 effects, the sponsor can evaluate whether the study results will be acceptable for registration on the basis of the accumulated evidence from the program; alternatively, whether the trial results could be used to define the inferential prior for a smaller follow up trial (Viele et al 2014).

### 5.3. *Considerations for the DMC and Interim Analyses*

For a trial with a DMC, the sponsor should ensure that the DMC is well-informed of all measures taken to protect participant safety and to address operational issues. Known or potential shortcomings of the data should be communicated. The timing of the regular pre-planned safety interim analyses may need to be re-assessed. In addition, revised or additional data presentations may be needed. In some circumstances of interim analysis discussed in this section, a DMC may need to be established if not already in existence. There could also be circumstances related to participant safety where there may be a need to urgently review unblinded data, and establishing an internal DMC that is appropriately firewalled from the rest of the study team is recommended (e.g., studies without an existing DMC where it could take many months to organize an external DMC).

Efficacy interim analyses should be conducted as planned with information level (e.g., number of participants with primary endpoint or specific information fraction) as described in the protocol, which may cause a delay in the expected timeline. Intermediate unplanned efficacy or futility interim analyses are generally discouraged unless there are safety and ethical considerations. However, if it is not feasible to reach the planned information level, altering the plan for interim analysis would need to be considered, for example with timing based on calendar time. In cases with strong scientific rationale for an unplanned interim analysis, the DMC should be informed and consulted on the time and logistics of the analysis. If an estimand, planned analysis methods, and/or decision rules have been changed to address pandemic-related disruptions and missing data, these changes should be communicated to the DMC and documented in the DMC charter.

We do not generally advocate utilizing a DMC for operational risk assessment / mitigation process, to prevent influence of unblinded data on trial conduct decisions (EMA/CHMP 2005, FDA 2006). If the sponsor decides to involve the DMC, details should be clearly defined and documented in the DMC charter, including additional responsibilities of DMC members and measures to prevent introduction of bias.

## 6. Conclusions

As we have discussed, the COVID-19 outbreak continues to have major impact on planned and ongoing clinical trials. The effects on trial data have multiple implications. In many cases these may go beyond the individual clinical trial and will need to be considered when such results are included with other trial results, such as an Integrated Summary of Efficacy or Safety. Our goal was to describe the nature of the statistical issues arising from COVID-19 potential impact on ongoing clinical trials and make general recommendations for solutions to address the issues.

The following are the most important findings and recommendations:
- Risk assessment, mitigation measures, and all changes to study conduct, data collection, and analysis must be documented in Statistical Analysis Plans and

Clinical Study Reports as appropriate. Some changes may necessitate protocol amendments and consultation with regulatory agencies (FDA 2020, EMA 2020a, 2020b).

- Implications of the operational mitigations for the statistical analysis of the trial data should be considered before implementing those mitigations, especially for key efficacy and safety endpoints. Validity and exchangeability of alternate methods of data collection require careful consideration.

- The estimand framework, comprised of five key attributes, provides a pathway for assessing the impact of the pandemic on key study objectives in a systematic and structured manner and may be useful regardless of whether estimands are formally defined in the study protocol.

- As much as possible, we recommend that original objectives of the trial be maintained; but some impact to planned estimands may be unavoidable. Pandemic-related intercurrent events will likely need to be defined to properly and rigorously account for unexpected pandemic effects.

- Planned efficacy and safety analyses should be reviewed carefully for changes needed to ensure that the estimators and missing data strategies align with updated estimands. Additional sensitivity and supportive analyses will be needed to fully describe the impact of the pandemic-related disruptions.

- Sponsors should make every effort to minimize missing data without compromising safety of participants and study personnel and without placing undue burden on the healthcare system. Priority should be on the assessments which determine the primary endpoint, important safety endpoints, followed by the key secondary efficacy endpoints.

- Most data that are missing due to pandemic reasons may be argued to be MCAR or MAR, especially if missingness is due to structural reasons, but additional considerations may apply, especially for certain diseases and participant-specific missingness.

- Sponsors should carry out rigorous and systematic risk assessment concerning trial and data integrity and update it regularly. The ability of trials to meet their objectives should be assessed quantitatively, taking account of the impacts on trial estimands, missing data and missing data handling, and needed modifications to analysis methods.

**Acknowledgements**

We are grateful for the help of colleagues at each of our companies who have devoted much time to addressing these issues in their ongoing clinical trials. They have generously shared their ideas, and this manuscript has benefited from this broad input. We also thank the members of the "Biopharmaceutical Statistics Leaders Consortium" who brought the team of authors together and provided valuable input; and the associate editor and reviewers who provided extensive and helpful input within tight timelines.

**References**

European Medicines Agency Committee For Medicinal Products For Human Use (EMA/CHMP) (2005), *Guideline On Data Monitoring Committees,* https://www.ema.europa.eu/en/documents/scientific-guideline/guideline-data-monitoring-committees_en.pdf

EMA/CHMP (2020a), *Guidance to Sponsors on How to Manage Clinical Trials During the COVID-19 Pandemic*, https://www.ema.europa.eu/en/documents/press-release/guidance-sponsors-how-manage-clinical-trials-during-covid-19-pandemic_en.pdf

EMA/CHMP (2020b), *Points to Consider on Implications of Coronavirus Disease (COVID-19) on Methodological Aspects of Ongoing Clinical Trials,* EMA/158330/2020. https://www.ema.europa.eu/en/documents/scientific-guideline/points-consider-implications-coronavirus-disease-covid-19-methodological-aspects-ongoing-clinical_en.pdf

U.S. Food and Drug Administration (FDA) (2006), *Guidance for Clinical Trial Sponsors: Establishment and Operation of Clinical Trial Data Monitoring Committees,* https://www.fda.gov/regulatory-information/search-fda-guidance-documents/establishment-and-operation-clinical-trial-data-monitoring-committees

FDA (2020), *Guidance on Conduct of Clinical Trials of Medical Products During COVID-19 Public Health Emergency*, Guidance for Industry, Investigators, and Institutional Review Boards. Updated on May 14, 2020. https://www.fda.gov/regulatory-information/search-fda-guidance-documents/fda-guidance-conduct-clinical-trials-medical-products-during-covid-19-pandemic

ICH (2019), *Addendum on Estimands and Sensitivity Analysis in Clinical Trials to The Guideline on Statistical Principles for Clinical Trials*, https://database.ich.org/sites/default/files/E9-R1_Step4_Guideline_2019_1203.pdf

ICH (2017), *General Principles for Planning and Design of Multiregional Clinical Trials*, https://database.ich.org/sites/default/files/E17EWG_Step4_2017_1116.pdf

Bogaerts, K., Komárek, A., and Lesaffre, E. (2017), *Survival Analysis with Interval-Censored Data: A Practical Approach*, ISBN 978-1-4200-7747-6, London: Chapman and Hall/CRC

Boyd, A.P., Kittelson, J.M., and Gillen, D.L. (2012), "Estimation of Treatment Effect Under Non-Proportional Hazards and Conditionally Independent Censoring," *Statistics in Medicine,* 31, 3504-3515.

Carpenter, J.R., Roger, J.H., and Kenward, M.G. (2013), "Analysis of Longitudinal Trials with Protocol Deviation: A Framework for Relevant, Accessible Assumptions, and Inference via Multiple Imputation," *Journal of Biopharmaceutical Statistics*, 23, 1352-1371.

Hughes, R. A., Sterne, J. A. C., and Tilling, K. (2016), "Comparison of Imputation Variance Estimators," *Stat. Methods Med. Res.,* 25, 2541–2557.

Kieser, M. and Friede, T. (2003), "Simple Procedures for Blinded Sample Size Adjustment That Do Not Affect the Type I Error Rate," *Statistics in Medicine,* 22, 3571–3581.

Mallinckrodt, C. H., Bell, J., Liu, G., Ratitch, B., O'Kelly, M., Lipkovich, I., Singh, P., Xu, L., and Molenberghs, G. (2019), "Aligning Estimators With Estimands in Clinical Trials: Putting the ICH E9(R1) Guidelines Into Practice," *Therapeutic Innovation & Regulatory Science*, doi.org/10.1177/2168479019836979

Mallinckrodt, C. H., Molenberghs G., Lipkovich I., and Ratitch B. (2020), *Estimands, Estimators, and Sensitivity Analysis in Clinical Trials*, London: Chapman & Hall/CRC

Mao, H., Li, L., Yang, W., and Shen, Y. (2018), "On the Propensity Score Weighting Analysis with Survival Outcome: Estimands, Estimation, and Inference," *Statistics in Medicine,* 37, 3745-3763.

Molenberghs, G. and Kenward, M. G. (2007), *Missing Data in Clinical Studies*, New York: John Wiley & Sons, Inc.

Little, R. J. A. and Rubin, D. B. (2002), *Statistical Analysis with Missing Data*, New York: John Wiley & Sons, Inc.

Müller, H.H. and Schäfer, H. (2004), "A General Statistical Principle for Changing a Design Any Time During the Course of a Trial," *Statistics in Medicine,* 23, 2497–2508.

Nguyen, V. Q. and Gillen, D. L. (2012), "Robust Inference in Discrete Hazard Models for Randomized Clinical Trials," *Lifetime Data. Anal*., 18, 446-69

National Research Council (US) Panel on Handling Missing Data in Clinical Trials (2010), *The Prevention and Treatment of Missing Data in Clinical Trials*, Washington (DC): National Academies Press (US).

O'Kelly, M. and Ratitch, B. (2014), *Clinical Trials with Missing Data: A Guide for Practitioners,* New York: John Wiley & Sons, Ltd.

Posch, M. and Proschan, M.A. (2012), "Unplanned Adaptations Before Breaking the Blind," *Statistics in Medicine,* 31, 4146–4153.

Ratitch, B., O'Kelly, M., and Tosiello, R. (2013), "Missing Data in Clinical Trials: from Clinical Assumptions to Statistical Analysis Using Pattern Mixture Models," *Pharmaceutical Statistics*, 12, 337-347.

Rubin, D.B. (1987), *Multiple Imputation for Nonresponse in Surveys*, New York: John Wiley & Sons, Inc.

Rufibach, K. (2019), "Treatment Effect Quantification for Time-to-event Endpoints – Estimands, Analysis Strategies, and Beyond," *Pharmaceutical Statistics*, 18, 145-165.

Schomaker, M. and Heumann, C. (2018), "Bootstrap Inference When Using Multiple Imputation," *Statistics in Medicine,* 37, 2252–2266.

Unkel, S., Amiri, M., Benda, N., et al. (2019), "On Estimands and the Analysis of Adverse Events in the Presence of Varying Follow-up Times Within the Benefit Assessment of Therapies," *Pharmaceutical Statistics*, 18, 165- 183. https://doi.org/10.1002/pst.1915

van Walraven, C., Davis, D., Forster, A.J., and Wells, G.A. (2004), "Time-dependent Bias was Common in Survival Analyses Published in Leading Clinical Journals," *Journal of Clinical Epidemiology*, 57, 672–82.

Viele, K., Berry, S., Neuenschwander, B., et al (2014), "Use of Historical Control Data for Assessing Treatment Effects in Clinical Trials," *Pharmaceutical Statistics,* 13, 41–54.

Wassmer, G and Brannath, W. (2016), *Group Sequential and Confirmatory Adaptive Designs in Clinical Trials,* Switzerland: Springer International Publishing AG.